\documentclass[twocolumn,prb,aps]{revtex4}
\usepackage[dvips]{graphicx}
\begin{document}
\title{Fourier transformation and response functions}



\begin{abstract} 
We improve on Fourier transforms (FT) between imaginary time $\tau$ and 
imaginary frequency $\omega_n$ used in certain quantum cluster approaches
using the Hirsch-Fye method.  The asymptotic behavior of the electron 
Green's function can be improved by using a ``sumrule'' boundary condition 
for a spline. For response functions a two-dimensional FT of a singular 
function is required. We show how this can be done efficiently by splitting 
off a one-dimensional part  containing the singularity and by performing
a semi-analytical FT for the remaining more innocent two-dimensional part. 

\end{abstract}

\author{O. Gunnarsson$^{(1)}$, G. Sangiovanni$^{(2)}$, A. Valli$^{(2)}$ and M. W.  Haverkort$^{(1)}$}   
\affiliation{
${}^1$Max-Planck-Institut f\"ur Festk\"orperforschung, D-70506 Stuttgart, Germany  \\
${}^2$Institut f\"ur Festk\"orperphysik, Technische Universit\"at Wien, Vienna, Austria
}

\maketitle

Quantum cluster cluster theories, such as the dynamical cluster 
approximation (DCA) or the cellular dynamical mean-field theory 
(CDMFT),\cite{Jarrell} make it possible to calculate dynamical 
quantities, e.g.,   electron Green's functions or response functions, 
for strongly correlated systems. These calculations, however, are 
numerically very demanding. In the Hirsch-Fye\cite{Hirsch} method 
for solving the resulting cluster problem, one has to switch 
between imaginary times $\tau$ and imaginary frequencies $\omega_n$, 
which requires Fourier transforms (FT). In this paper we address 
efficient methods for performing FT in this context. In the weak-coupling 
version\cite{Rubtsov,CT,Assaad} of continuous time 
approaches\cite{Rubtsov,CT,Assaad,CTAUX} properties 
can be very easily measured directly in frequency space and no FT is needed. 
 
The FT of the electron Green's function has to be performed with 
great care, and it is important to obtain the correct asymptotic 
behavior for large $\omega_n$.\cite{Blumer,Comanac,Knecht,Gull}  
Inaccuracies for large $\omega_n$ can lead to problems, for instance 
in the DMFT description of a Mott transition.\cite{Blumer} The 
large $\omega_n$  behavior can be greatly improved if exact moments 
are calculated. The FT can then be performed using a natural spline, 
which works            well for a symmetric half-filled Hubbard 
model.\cite{Blumer} For a nonsymmetric or doped Hubbard model, 
however, the accuracy is not optimum. Here we show how this can be 
improved by introducing  ``sumrule'' boundary conditions for the 
spline interpolation. 

The main part of this paper deals with response functions.
This is important not only for studying susceptibilities but also
for diagrammatic extensions of the dynamical mean-field theory, such as 
the dynamical vertex approximation\cite{dynamical} and the dual
fermion method.\cite{dual} The most convenient way of of calculating 
them within DCA is to work directly in $\omega_n$-space. Yet, 
in the Hirsch-Fye method this requires FT for functions 
$g(\tau_i,\tau_j)$ depending on two imaginary times,
$\tau_i$ and $\tau_j$, for each Monte-Carlo step. The FT is 
complicated by the fact that these functions have singularities 
and that the precise asymptotic behavior is difficult to work out. 
We show that a very efficient solution is to split up 
$g(\tau_i,\tau_j)$ into two parts. 
One part, $g^{(0)}(\tau_i-\tau_j)$, depends only on the difference  
$\tau_i-\tau_j$ and contains the singularity.  The second part, 
$\delta g(\tau_i,\tau_j)$ depends on both variables independently 
but is well behaved. To perform spline interpolations in both 
variables is very time consuming. We show how the FT can be performed 
in a very efficient way for $\delta g(\tau_i,\tau_j)$.

If $G(\tau)$ is only known for discrete values of $\tau$ separated by 
$\Delta \tau$, a direct FT can only give accurate results up to 
$\omega_n \sim 1/\Delta \tau$. To improve the accuracy for large
$\omega_n$, on can subtract a model Green's function, $G_m(\tau)$, from  
$G(\tau)$, where $G_m$ has the right asymptotic behavior. The difference, 
$G(\tau)-G_m(\tau)$, is then FT and $G_m(i\omega_n)$ is added. For instance, 
$G_m(\tau)$ can be obtained from perturbation theory.\cite{Freericks} 
Alternatively, one can calculate the lowest moments of the spectral 
function exactly from appropriate expectation values.\cite{Blumer} 
$G_m(i\omega_n)$ can be chosen so that these moments are exactly satisfied, 
meaning that the corresponding coefficients in a $(1/\omega_n)$ expansion 
of $G_m(i\omega_n)$ are correct. $G(\tau)-G_m(\tau)$ is then  FT using a 
natural spline.\cite{Blumer} If the  FT of $G(\tau)-G_m(\tau)$ 
correctly gives the lowest moments equal to zero, it follows that 
the corresponding moments of $G(i\omega_n)$ are correct. 

To illustrate this, we consider the relation between the spectral function,
$A(\omega)$, and $G(\tau)$
\begin{equation}\label{eq:1}
G(\tau)=\int_{-\infty}^{\infty}{e^{-\omega \tau}\over 
1+e^{-\omega \beta}}A(\omega)d\omega,
\end{equation}
where $\beta=1/T$ and $T$ is the temperature.
This gives the following, very useful, relations 
\begin{equation}\label{eq:2}
G^{(n)}(0)+G^{(n)}(\beta)=\int_{-\infty}^{\infty}(- \omega)^n A(\omega)d\omega
\equiv (-1)^n M_n,
\end{equation}
where $G^{(n)}(\tau)=d^nG(\tau)/d \tau^n$ and $M_n$ is the $n$th moment. 
For large $\omega_n$, $G(\omega_n) \sim \sum_k M_k/(i\omega_n)^{k+1}$.
For the difference 
$\Delta G(\tau)\equiv G(\tau)-G_m(\tau)$, it then follows that
\begin{equation}\label{eq:3}
\Delta G^{(n)}(0)+\Delta G^{(n)}(\beta)=0 \hskip1cm n=0, 1, 2,
\end{equation}
if  $G_m$ has  the correct 0th, 1st and 2nd moments.  

Let $\Delta G(\tau)$ be given for $n$ 
$\tau$-values, $\tau_1$, ..., $\tau_n$. In a third order spline, 
each interval between two $\tau$-values is 
interpolated with a third order polynomial.  It is required that 
the polynomials give the correct $\Delta G(\tau_i)$
and that the first two derivatives are continuous at each $\tau_i$.
There are then $4(n-1)$ unknown coefficients 
and $4(n-2)+2$ conditions. We then have to provide two more
conditions.

For a natural spline,  it is assumed that $\Delta G^{(2)}(0)=
\Delta G^{(2)}(\beta)=0$, while the first derivatives are 
left open. The  assumption about $\Delta G^{(2)}$ is too
strong, since we only know that $\Delta G^{(2)}(0)+
\Delta G^{(2)}(\beta)=0$. The second moment is nevertheless 
correct, since it only depends on the sum. The incorrect 
assumption about $\Delta G^{(2)}$, however, influences the 
estimates of $\Delta G^{(1)}$ and $\Delta G^{(1)}(0)+\Delta G^{(1)}
(\beta)=0$ is in general not satisfied. This leads 
to incorrect results already for the important first moment.
Due to symmetry, the natural spline may give $\Delta G^{(1)}(0)
+ \Delta G^{(1)}(\beta)=0$ in special cases, e.g., at half-filling
for the symmetric Hubbard model. 

A better approach is to use the two conditions of Eq.~(\ref{eq:3})
for $n=1$ and 2.
We refer to this as the sumrule spline.
This gives correct first and second moments and large $\omega_n$ 
behavior, even if the estimates of $\Delta G^{(n)}(0)$ and 
$\Delta G^{(n)}(\beta)$ individually are not accurate. 

\begin{figure}
{\rotatebox{-90}{\resizebox{5.cm}{!}{\includegraphics {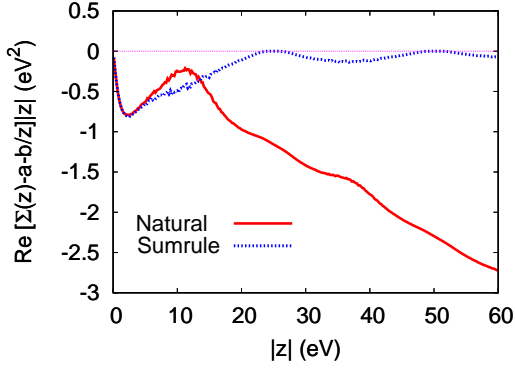}}}}
\caption{\label{fig:1}(color on-line) Deviation of the local 
$\Sigma(z)$ ($z=i\omega_n$) from the asymptotic form $\Sigma(z)\sim a+b/z$ 
for natural and sumrule  boundary conditions. For natural 
boundary conditions the frequency independent part $a$ has an error.                  
}  
\end{figure}

Fig.\ref{fig:1} compares the two approaches for the two-dimensional
(2d) Hubbard model in the DCA. We have used $t=-0.4$ and $t_p=-0.3t$  for the nearest 
and next nearest neighbor hopping, respectively, $U=8|t|$ for the on-site 
Coulomb interaction and $\beta |t|=12$. All energies
are in eV. The occupancy is $n=0.9$ and we considered an eight site cluster. 
The number of $\tau$-points is $N_{\tau}=120$. 
For large $\omega_n$, the local Green's function behaves as 
$G(z)\sim 1/(z-a-b/z)$, where $z=i\omega_n$ and  
$a=M_1$ and $b=M_2-M_1^2$ are given by the first two moments. 
If we define $\Sigma(z)=z-G^{-1}(z)$, we have 
\begin{equation}\label{eq:4b}
\Sigma(z)\sim a+b/z          
\end{equation}
for large $|z|$. The figure shows that Re $\Sigma(z)\to a$ is obtained 
for the sumrule spline  but not for the natural spline. 
Also for Im $\Sigma(z)\sim b/z$ (not shown in the figure) the natural spline is
less accurate than the sumrule spline, since an error in $M_1$ also 
enters in $b$, but for large $|z|$ the error for Im $\Sigma(z)$ is much
smaller than for Re $\Sigma(z)$.
As the write up of this work was being finished, we became aware of   
very similar approaches in the thesises by Comanac\cite{Comanac} and 
by  Gull.\cite{Gull}

We now discuss the FT for response functions, and use DCA as an illustration.
The response function is calculated for a cluster in a bath. 
From the cluster response function vertex corrections are deduced. 
The Bethe-Salpeter equation for the lattice is then solved, assuming 
that the vertex corrections are the same as for the cluster. 
The embedded cluster problem is solved for imaginary time, while the 
Bethe-Salpeter equation is solved for imaginary frequency. The necessary 
FT to imaginary frequencies is numerically difficult.

We consider the electron-hole response function
\begin{eqnarray}\label{eq:5}
&&\chi_{\sigma \sigma^{'}}(q,k,k^{'})=-{1\over \beta^2}\int_0^\beta d\tau_1
\int_0^\beta d\tau_2 \int_0^\beta d\tau_3 \int_0^\beta d\tau_4  \nonumber \\ 
&& \times {\rm exp}  \lbrace -i[\omega_n \tau_1-(\omega_n+\nu)\tau_2
+(\omega_{n^{'}}+\nu)\tau_3 -\omega_{n^{'}}\tau_4] \rbrace \nonumber          \\ 
&&\times \langle T_{\tau} [c^{\dagger}_{ {\bf k} \sigma}(\tau_1)c_{{\bf k+q} \sigma}(\tau_2)
c^{\dagger}_{{\bf k^{'}+q} \sigma^{'}}(\tau_3)c_{{\bf k^{'}}\sigma^{'}}
(\tau_4)]\rangle, 
\end{eqnarray}
with the compact notations $k=({\bf k}, \omega_n)$ and 
$q=({\bf q}, \nu)$. Here $T_\tau$
is a time ordering symbol, $c^{\dagger}_{{\bf k} \sigma}$ creates an
electron with wave vector ${\bf k}$ and spin $\sigma$ and                        
$c(\tau)={\rm exp}(H\tau)c{\rm exp}(-H\tau)$, where $H$ is the Hamiltonian.

The $c$ operators are contracted pairwise and their expectation 
values are calculated.  This is done for a very large number of 
configurations. It is convenient to perform the FT to imaginary frequency
and reciprocal space for each configuration and to store the results in the
 $k$- and $q$-variables,\cite{Jarrell} rather than storing the results 
in imaginary time and real space. 
We then need  
\begin{eqnarray}\label{eq:6}
&&g_{\sigma}(k,k^{'})=\sum_{i=1}^{N_{\tau}}\sum_{j=1}^{N_{\tau}} 
\sum_{{\bf R}_i {\bf R}_j} e^{   -i[\omega_n \tau_i
-\omega_{n^{'}}\tau_j] }  \\ && \times e^{  i[{\bf k}
\cdot {\bf R}_i-{\bf k^{'}}\cdot {\bf R}_j]} 
\langle T_{\tau}[c^{\phantom \dagger}_{{\bf R}_j \sigma}(\tau_j) 
c^{\dagger}_{ {\bf R}_i \sigma}(\tau_i)]\rangle (\Delta \tau)^2,
\nonumber
\end{eqnarray}
where the integrals over $\tau$ have been replaced by sums over $N_{\tau}$ 
discrete values of $\tau$ separated by $\Delta \tau$ and ${\bf R}$ is a site index.
Although this has the form of a Green's function, it is calculated
for a specific configuration and only the zeroth moment is known.               
This makes it harder to perform a FT. There is a singularity 
at $\tau_i=\tau_j$, which makes a straightforward spline in  
$\tau_i$ and $\tau_j$ less useful. The singularity can be handled
by treating $\tau_i <  \tau_j$ and $\tau_i >  \tau_j$ separately.
But a spline in two variables is still numerically very demanding because  
of the large number of points needed. To see this, we simplify the  
calculation in Eq.~(\ref{eq:6}), by splitting it in two parts.
Thus we calculate 
\begin{eqnarray}\label{eq:7}
&& f_{\sigma}(k,{\bf R}_j,\tau_j)=\sum_{i, {\bf R}_i}  
e^{-i(\omega_n \tau_i -{\bf k}
\cdot {\bf R}_i) } \\ &&   \times    \langle T_{\tau}[ c_{{\bf R}_j \sigma}(\tau_j) 
c^{\dagger}_{ {\bf R}_i \sigma}(\tau_i))]\rangle,
\nonumber
\end{eqnarray}
and
\begin{eqnarray}\label{eq:8}
&&g_{\sigma}(k,k^{'})=\sum_{j,{\bf R}_j}   
e^{ i(\omega_{n^{'}} \tau_j 
-{\bf k^{'}}\cdot {\bf R}_j) }       f_{\sigma}
(k,{\bf R}_j,\tau_j).
\end{eqnarray}
This gives
\begin{eqnarray}\label{eq:9}
&&\chi_{\sigma \sigma^{'}}(q,k,k^{'})
=-{1\over \beta^2}[g_{\sigma}
(k+q,k)g_{\sigma^{'}}(k^{'},k^{'}+q)\nonumber \\
&&-g_{\sigma}(k^{'},k)
g_{\sigma}(k+q,k^{'}+q)\delta_{\sigma \sigma^{'}}](\Delta \tau)^2
\end{eqnarray}
Let the number of sites be $N_c$, the number of $\omega_n$-values 
$N_{\omega}$ and the number of $\nu$-values $N_{\nu}$.  The number 
of ${\bf k}$-values is then also $N_c$. Let $N_{\tau}^{\rm c}$ be the
number of $\tau$-points for which the correlation function in 
Eq.~(\ref{eq:7}) is known and $N_{\tau}^{\rm s}$ the number of 
$\tau$-values needed to obtain an accurate FT.  
Eq.~(\ref{eq:7}) and Eq.~(\ref{eq:8}) then require of the order of 
$2(N_{\omega}+N_{\nu})N_{\tau}^{\rm c}N_{\tau}^{\rm s}N_c^3$ and 
$2(N_{\omega}+N_{\nu})^2N_{\tau}^{\rm s}N_c^3$ operations, respectively, 
for each configuration. Here we have assumed that the spline in the
first $\tau$-variable is only done for each of the $N_{\tau}^{\rm c}$
values of the second $\tau$-variable. After the corresponding FT has
been performed, the second variable is splined and FT. The calculations 
can easily be arranged so that the time needed for calculating the 
exponents is negligible and efficient machine routines can be used for
the multiplications. Still, the calculations are very 
time consuming if $N_{\tau}^{\rm s}$ is large enough to give accurate 
FT. We therefore follow a different route, reducing     the time 
requirement for Eqs~(\ref{eq:7}, \ref{eq:8}) very substantially and 
requiring     no interpolation of the $\tau$-variables. 

We first notice that
\begin{equation}\label{eq:10}
g_{{\sigma \bf R}_i,{\bf R}_j}(\tau_i, \tau_j)=\langle[T_{\tau} 
c^{\phantom \dagger}_{{\bf R}_j \sigma}(\tau_j)
c^{\dagger}_{ {\bf R}_i \sigma}(\tau_i))]\rangle
\end{equation}
depends on $\tau_i$ and $\tau_j$ individually and not only on their 
difference, since it is calculated for one particular configuration. 
However, we can separate it as
\begin{equation}\label{eq:11}
g_{{\sigma \bf R}_i {\bf R}_j}(\tau_i, \tau_j)\equiv g^{0}_{{\sigma \bf R}_i {\bf R}_j}(\tau_i-\tau_j)
+\delta g_{{\sigma \bf R}_i {\bf R}_j}(\tau_i, \tau_j),
\end{equation}
where
\begin{equation}\label{eq:12}
g^{0}_{{\sigma \bf R}_i {\bf R}_j}(\tau_i)={1\over N_{\tau}}\sum_{j=1}^{N_{\tau}}
g_{{\sigma \bf R}_i {\bf R}_j}(\tau_{i+j-1}, \tau_j), 
\end{equation}
only depends on one $\tau$-variable and we have defined 
$g_{{\sigma \bf R}_i {\bf R}_j}(\tau_{i+j-1}, \tau_j)=-g_{{\sigma \bf R}_i {\bf R}_j}
(\tau_{i+j-1-N_{\tau}}, \tau_j)$ if $i+j-1>N_{\tau}$ or $g_{{\sigma \bf R}_i {\bf R}_j}(\tau_{i+j-1}, \tau_j)=-g_{{\sigma \bf R}_i {\bf R}_j}
(\tau_{i+j-1+N_{\tau}}, \tau_j)$
$i+j-1<1$. Here, $g^{0}_{{\sigma \bf R}_i {\bf R}_j}(\tau_i)$ 
is not a noninteracting  Green's function, but the time
translationally invariant part of $g_{{\sigma \bf R}_i {\bf R}_j}(\tau_{i+j-1}, \tau_j)$. 
The singularities are now in  $g^{0}_{{\sigma \bf R}_i {\bf R}_j}(\tau)$, 
and  $\delta g_{{\sigma \bf R}_i {\bf R}_j}(\tau_i,\tau_j)$ 
is free of singularities, and can more easily be Fourier transformed.

Since $g^{0}_{{\sigma \bf R}_i {\bf R}_j}(\tau)$ only depends on one variable, it
can easily be FT using a spline. Alternatively, we can use Filon's 
rule,\cite{Abramowitz} where second order polynomials are 
fitted to the $N_{\tau}$ $\tau$-points. 
These polynomials are then FT analytically.  Even for  $\omega_n \Delta 
\tau\gg 1$, the FT can be  very accurate. This automatically gives 
the appropriate $1/(i \omega_n)$ behavior for large $\omega_n$, due to 
end point corrections.

It is possible to FT $\delta g_{{\sigma \bf R}_i {\bf R}_j}(\tau_i,\tau_j)$
by performing a Filon's rule for first $\tau_i$ and then for $\tau_j$.
However, we have found it preferable to fit a two-dimensional
polynomial  
\begin{equation}\label{eq:13}
a_{00}+a_{10}\tau+a_{01}\tau'+a_{11}\tau\tau',
\end{equation} 
to the values of $\delta g_{{\sigma \bf R}_i {\bf R}_j}(\tau,\tau')$ in the points
$(\tau_i,\tau_j)$, $(\tau_{i+1},\tau_j)$, $(\tau_{i},\tau_{j+1})$ and 
$(\tau_{i+1},\tau_{j+1})$. This is multiplied by the appropriate
exponent and integrated analytically. Substantial simplification follow
from the fact that $\delta g_{{\sigma \bf R}_i {\bf R}_j}(\tau,\tau^{'})$ 
is antiperiodic and exp$[i(\omega_n \tau-\omega_{n^{'}}\tau']
\delta g_{{\sigma \bf R}_i {\bf R}_j}(\tau,\tau^{'})$
is periodic in $\tau$ and $\tau'$. Then
\begin{eqnarray}\label{eq:14}
&&\int_0^{\beta}d\tau \int_0^{\beta}d\tau' e^{i[\omega_n \tau
-\omega_{n^{'}}\tau']}\delta g_{{\sigma \bf R}_i {\bf R}_j}(\tau,\tau^{'})=\\
&&{1\over 4}c(\omega_n,\omega_{n^{'}},\Delta \tau/2)  
\sum_{i,j=1}^{N_{\tau}} e^{i[\omega_n \tau_i
-\omega_{n^{'}}\tau_j]}\delta g_{{\sigma\bf R}_i {\bf R}_j}(\tau_i,\tau_j),
\nonumber
\end{eqnarray} 
where $\tau_i=(i-1)\Delta \tau$ and $(N_{\tau}+1)\Delta \tau=\beta$.
Here
\begin{eqnarray}\label{eq:14a}
&&c(x,y,\Delta)=  \nonumber \\
&&e^{-i\Delta (x-y)} [b_0^{x}b_0^{-y}+b_0^{x}b_1^{-y}
+b_1^{x}b_0^{-y} +b_1^{x}b_1^{-y}] \nonumber \\
&&e^{-i\Delta (x+y)} [b_0^{x}b_0^{-y}-b_0^{x}b_1^{-y}
+b_1^{x}b_0^{-y} -b_1^{x}b_1^{-y}] \nonumber \\
&&e^{i\Delta (x+y)} [b_0^{x}b_0^{-y}+b_0^{x}b_1^{-y}
-b_1^{x}b_0^{-y} -b_1^{x}b_1^{-y}] \nonumber \\
&&e^{i\Delta (x-y)} [b_0^{x}b_0^{-y}-b_0^{x}b_1^{-y}
-b_1^{x}b_0^{-y} +b_1^{x}b_1^{-y}] \nonumber  
\end{eqnarray}
where 
\begin{eqnarray}\label{eq:14b}
&&b_0^x=-{i \over x}(e^{ix\Delta}-e^{-ix \Delta}) \\ \nonumber
&&b_1^x={1\over x^2\Delta}[e^{ix\Delta}(1-ix\Delta)-e^{-ix\Delta}(1+ix\Delta)]
\end{eqnarray}
This approach can easily be extended to the case of a nonuniform grid.

If $\delta g_{{\sigma \bf R}_i {\bf R}_j}(\tau,\tau^{'})$ were a 
very smooth function, a more accurate integration method could be 
devised by fitting a polynomial of higher order. However, since 
$\delta g$ is obtained for a specific configuration, this does 
not seem useful.       

\begin{figure}
{\rotatebox{0}{\resizebox{6.5cm}{!}{\includegraphics {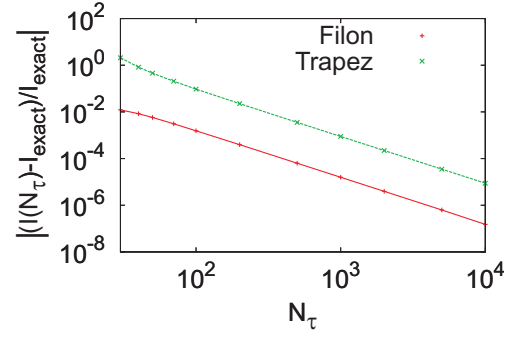}}}}
\caption{\label{fig:2}(color on-line) The relative accuracy of 
the FT of Eq.~(\ref{eq:14c}) according to the approach of Eq.~(\ref{eq:14}) 
(Filon) or using the trapezoidal rule (Trapez). 
}  
\end{figure}

To test the method,  we have FT a function
\begin{equation}\label{eq:14c}
f(\tau_1,\tau_2)=\sum_{ij}a_{ij}({\tau_1-\beta/2 \over \beta/2})^i
({\tau_2-\beta/2 \over \beta/2})^j,
\end{equation}
where $a_{ij}$ is only nonzero for odd values of $i$ and $j$ to assure that
the function is antiperiodic. Specifically, we chose $a_{11}=0.7$,
$a_{13}=1.3$, $a_{15}=0.9$, $a_{31}=-1.2$, $a_{33}=1.5$, $a_{35}=-0.6$,
$a_{51}=-0.8$, $a_{53}=1.1$, $a_{55}=-0.7$. We used the frequencies
$\omega_n=13.5(2\pi/\beta)$ and $\omega_n^{'}=9.5(2\pi/\beta)$, where
$\beta=15$.  Fig.~\ref{fig:2} shows results obtained by using Eq.~(\ref{eq:14}) (Filon)
or the simple trapezoidal rule (Trapez). In the figure, the approach of Filon 
leads to a comparable accuracy as the trapezoidal rule for a $N_{\tau}$ that is  
almost one order of magnitude smaller. 

In this Filon like approach the exponent is treated exactly and the error
in the FT is entirely due to the limited information about the function
to be FT. It is then no gain in adding extra points by interpolating the
function to be FT. In a Hirsch-Fye approach this means that we put
$N_{\tau}=N_{\tau}^{\rm c}$, the number of points determined by
the discretization used. 

From $\chi_{\sigma \sigma^{'}i}(q,k,k^{'})$ [Eq.~(\ref{eq:5})] we can calculate 
$\Pi(\tau)=\langle {\bf j}(\tau)\cdot {\bf j}(0)\rangle/(3N)$,
where $N$ is the number of lattice sites and ${\bf j}$ is the 
current operator. The FT of $\Pi(\tau)$ is related to 
the optical conductivity $\sigma(\omega)$ via
\begin{equation}\label{eq:16}
\Pi(\nu)=
 {1\over \pi}\int_{-\infty}^{\infty} {\omega^2 \over \nu^2+\omega^2}\sigma(\omega)d\omega.
\end{equation}

\begin{figure}
{\rotatebox{-90}{\resizebox{5.cm}{!}{\includegraphics {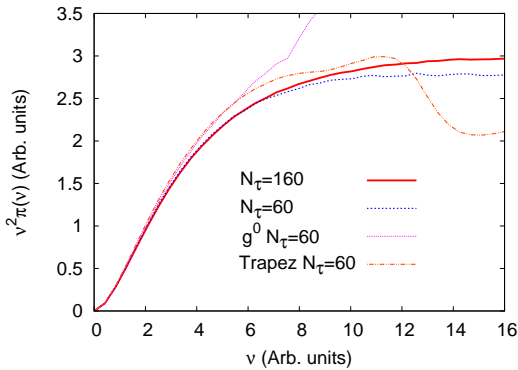}}}}
\caption{\label{fig:3}(color on-line)The quantity $\nu^2\Pi(\nu)$
as a function of $\nu$ for different values of $N_{\tau}$.  
The figure also shows results when $g^0$ has been split off [Eq.~(\ref{eq:11})]
but the trapezoidal rule was used for $\delta g$ ($g^0$) 
or the total $g$ was integrated  using a trapezoidal rule (Trapez).
}  
\end{figure}
Eq.~(\ref{eq:16}) shows that $\nu^2\Pi(\nu)$ approaches a constant 
for large $\nu$. Problems of the FT should show up in particular for 
large $\nu$ and the accuracy should increase with $N_{\tau}$. 
We then choose $N_{\tau}$ so large that $\nu^2\Pi(\nu)$
is constant for large values of $\nu$-values. This should then be an 
accurate result. 

Fig.~\ref{fig:3} shows results for $\nu^2\Pi(\nu)$ for the 2d 
Hubbard model. The parameters are the same as in Fig.~\ref{fig:1}, 
except that $\beta=15$. The bath obtained for $N_{\tau}=160$ was 
used also for $N_{\tau}=60$. For $N_{\tau}=160$, $\nu^2\Pi(\nu)$ 
is constant for large $\nu$ over the whole range shown.  The comparison 
with $N_{\tau}=60$ suggests that the FT is quite accurate at least 
for $\nu \Delta \tau \lesssim 2$ and it stays fairly accurate for substantially 
larger values $\nu \Delta \tau$. The deviation between $N_{\tau}
=60$ and 160 could also be due to other inaccuracies for $N_{\tau}=60$ 
than the FT, and in that case the FT is accurate for even larger $\nu \Delta \tau$.  
The figure  also shows a calculation where we split off $g^0$,
[Eq.~(\ref{eq:11})], and FT it using Filon's rule, but FT $\delta g$
using the trapezoidal rule ($g^0$ in the figure). We also performed
the FT on the full $g$, without splitting off $g^0$, using the trapezoidal
rule (Trapez in the figure). The figure shows that for  
$N_{\tau}^{\rm c}=60$ both approaches fail dramatically for large $\nu$.

To summarize, the FT of the Green's function can be improved by using a 
spline with sumrule boundary conditions. This gives a $G(i\omega_n)$
with correct first and second moments, while a natural spline in general
gives an incorrect first moment. To calculate a response function, we need a FT
a function $g(\tau_i,\tau_j)$ with a singularity. We show how a $g^0$ can be split
off, which only depends on the difference $\tau_i-\tau_j$ and which contains
the singularity. This function can be FT very accurately. For the rest, $\delta g$, 
we developed a two-dimensional FT in the spirit of Filon's rule. This leads to
accurate results, even if $g(\tau_i,\tau_j)$ is only known on a rather sparse mesh.
"After this paper had been submitted, an alternative prescription for
efficient Fourier transforms of two-particle Green's functions has been proposed
by Kunes.\cite{Kunes}

We would like to thank  E. Gull, F. Assaad and A. Toschi for useful 
discussions and J. Bauer for a careful reading of the manuscript. 
One of us (G.S.) acknowledges support from the FWF under “Lise-Meitner”
Grant No. M1136

\end{document}